\documentclass{emulateapj}

\newcommand{\Dnu}{\mbox{$\Delta \nu$}}
\newcommand{\acena}{\mbox{$\alpha$~Cen~A}}
\newcommand{\acenb}{\mbox{$\alpha$~Cen~B}}

\newcommand{\bhyi}{\mbox{$\beta$~Hyi}}

\newcommand{\nuind}{\mbox{$\nu$~Ind}}

\newcommand{\cms}{\mbox{cm\,s$^{-1}$}}

\newcommand{\ms}{\mbox{m\,s$^{-1}$}}
\newcommand{\muHz}{\mbox{$\mu$Hz}}

\newcommand{\newnew}[1]{\relax #1}
\newcommand{\half}{{\textstyle\frac{1}{2}}}

\slugcomment{to appear in ApJ}

\shorttitle{Oscillations in $\beta$ Hydri}
\shortauthors{Bedding et al.}

\begin{document}

\title{Solar-like oscillations in the G2 subgiant $\beta$ Hydri from
  dual-site observations}

\author{
Timothy~R.~Bedding,\altaffilmark{1}
Hans~Kjeldsen,\altaffilmark{2}
Torben Arentoft,\altaffilmark{2}
Francois~Bouchy,\altaffilmark{3,4}
Jacob~Brandbyge,\altaffilmark{2}
Brendon~J.~Brewer,\altaffilmark{1}
R.~Paul~Butler,\altaffilmark{5}
J{\o}rgen~Christensen-Dalsgaard,\altaffilmark{2}
Thomas~Dall,\altaffilmark{6}
S{\o}ren~Frandsen,\altaffilmark{2}
Christoffer~Karoff,\altaffilmark{2}
L\'aszl\'o~L.~Kiss,\altaffilmark{1}
Mario~J.P.F.G.~Monteiro,\altaffilmark{7}
Frank~P.~Pijpers,\altaffilmark{8}
Teresa~C.~Teixeira,\altaffilmark{7}
C.~G.~Tinney,\altaffilmark{9}
Ivan~K.~Baldry,\altaffilmark{9,10}
Fabien~Carrier,\altaffilmark{3,11} and
Simon~J.~O'Toole\altaffilmark{1,9}
}

\altaffiltext{1}{School of Physics A28, University of Sydney, NSW 2006,
Australia; bedding@physics.usyd.edu.au; brewer@physics.usyd.edu.au; laszlo@physics.usyd.edu.au}

\altaffiltext{2}{Danish AsteroSeismology Centre (DASC), Department of
Physics and Astronomy, University of Aarhus, DK-8000 Aarhus C, Denmark;
hans@phys.au.dk, toar@phys.au.dk, jacobb@phys.au.dk, srf@phys.au.dk, jcd@phys.au.dk,
karoff@phys.au.dk}

\altaffiltext{3}{Observatoire de Gen\`eve, Ch.~des Maillettes 51, CH-1290
Sauverny, Switzerland; francois.bouchy@obs.unige.ch}

\altaffiltext{4}{Laboratoire d'Astrophysique de Marseille, Traverse du
Siphon, BP 8, 13376 Marseille Cedex 12, France}

\altaffiltext{5}{Carnegie Institution of Washington,
Department of Terrestrial Magnetism, 5241 Broad Branch Road NW, Washington,
DC 20015-1305; paul@dtm.ciw.edu}

\altaffiltext{6}{European Southern Observatory, Casilla 19001, Santiago 19,
Chile; tdall@eso.org}

\altaffiltext{7}{Faculdade de Ci\^encias and Centro deCentro de
Astrof{\'\i}sica da Universidade do Porto, Rua das Estrelas, 4150-762
Porto, Portugal; mjm@astro.up.pt}

\altaffiltext{8}{Blackett Laboratory, Imperial College London, South
Kensington, London SW7 2BW, UK; f.pijpers@imperial.ac.uk}

\altaffiltext{9}{Anglo-Australian Observatory, P.O.\,Box 296, Epping, NSW
1710, Australia; cgt@aaoepp.aao.gov.au; otoole@aaoepp.aao.gov.au}

\altaffiltext{10}{Astrophysics Research Institute, Liverpool John Moores
University, Egerton Wharf, Birkenhead, CH41 1LD, UK; ikb@astro.livjm.ac.uk}

\altaffiltext{11}{Instituut voor Sterrenkunde, Katholieke Universiteit
Leuven, Celestijnenlaan 200 B, 3001 Leuven, Belgium;
fabien@ster.kuleuven.be}

\begin{abstract} 
We have observed oscillations in the nearby G2 subgiant star \bhyi{} using
high-precision velocity observations obtained over more than a week with
the HARPS and UCLES spectrographs.  The oscillation frequencies show a
regular comb structure, as expected for solar-like oscillations, but with
several $l=1$ modes being strongly affected by avoided crossings.  The
data, combined with those we obtained five years earlier, allow us to
identify 28 oscillation modes.  By scaling the large frequency separation
from the Sun, we measure the mean density of \bhyi{} to an accuracy of
0.6\%.  The amplitudes of the oscillations are about 2.5 times solar and
the mode lifetime is 2.3\,d.  A detailed comparison of the mixed $l=1$
modes with theoretical models should allow a precise estimate of the age of
the star.
\end{abstract}


\section{Introduction}

There has been tremendous recent progress in observing solar-like
oscillations in main-sequence and subgiant stars.  In a few short years we
have moved from ambiguous detections to firm measurements (see
\citealt{B+K2006b} for a recent summary).  Most of the results have come
from high-precision Doppler measurements using spectrographs such as
CORALIE, HARPS, UCLES and UVES\@.

The star $\beta$~Hydri (HR 98, HD 2151, HIP 2021) is a bright southern G2
subgiant ($V=2.80$) that is an excellent target for asteroseismology.
\newnew{Early unsuccessful attempts to measure oscillations were made by
\citet{Fra87} and \citet{E+C95}.}  Five years ago we reported oscillations
in this star from velocity measurements made with UCLES \citep{BBK2001} and
confirmed with CORALIE \citep{CBK2001}.  Those observations implied a large
frequency separation of 56--58\,\muHz{} but did not allow the unambiguous
identification of individual modes.  Meanwhile, theoretical models for
\bhyi{} have been published by \citet{F+M2003} and \citet{DiMChDP2003},
with both studies indicating the occurrence of avoided crossings for modes
with $l=1$ (see \S\,\ref{sec.freqs}).  This goes a long way toward
explaining the earlier difficulty in mode identification.

Here we report new observations of \bhyi{} using HARPS and UCLES in a
dual-site campaign.  We confirm the earlier detection of oscillations and
are able to identify nearly 30 modes, including some which show the clear
effect of avoided crossings.  We also measure the amplitudes of the
oscillations and the mode lifetimes, and use the large separation to infer
the mean stellar density.

\section{Velocity observations and power spectra}

We observed \bhyi{} in 2005 September from two sites.  
At the European Southern Observatory on La Silla in Chile we used the HARPS
spectrograph (High Accuracy Radial velocity Planet Searcher) with the 3.6-m
telescope\footnote{Based on observations collected at the European Southern
Observatory, La Silla, Chile (ESO Programme 75.D-0760(A)).}\@.  A thorium
emission lamp was used to calibrate the velocities, which were extracted
using the HARPS pipeline \citep{RPM2004}.
At Siding Spring Observatory in Australia we used UCLES (University College
London \'Echelle Spectrograph) with the 3.9-m Anglo-Australian Telescope
(AAT).  An iodine absorption cell was used to provide a stable wavelength
reference, with the same setup that we have previously used with this
spectrograph for \bhyi{} \citep{BBK2001} and other stars.

With HARPS we obtained 2766 spectra of \bhyi, with a dead time between
exposures of 31\,s and exposure times of 40 or 50\,s, depending on the
conditions.
With UCLES we obtained 1191 spectra, with a dead time between exposures of
70\,s and exposure times in the range 60--180\,s, such that the median
sampling time was 187\,s (implying a Nyquist frequency of about 2.7\,mHz).

\begin{figure*}
\epsscale{0.9}
\plotone{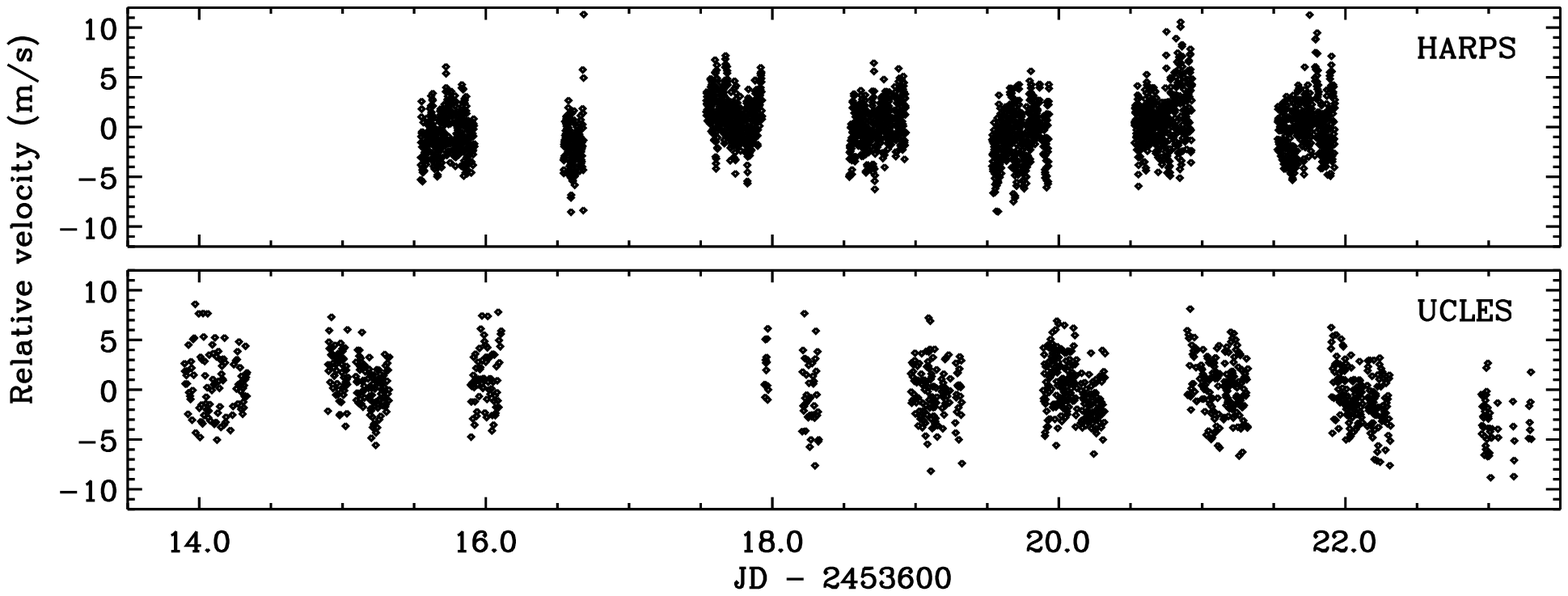}
\caption[]{\label{fig.series} Time series of velocity measurements of
\bhyi{} obtained from 2005 August 31 to September 9.  }
\end{figure*}

\begin{figure}
\epsscale{0.45}
\plotone{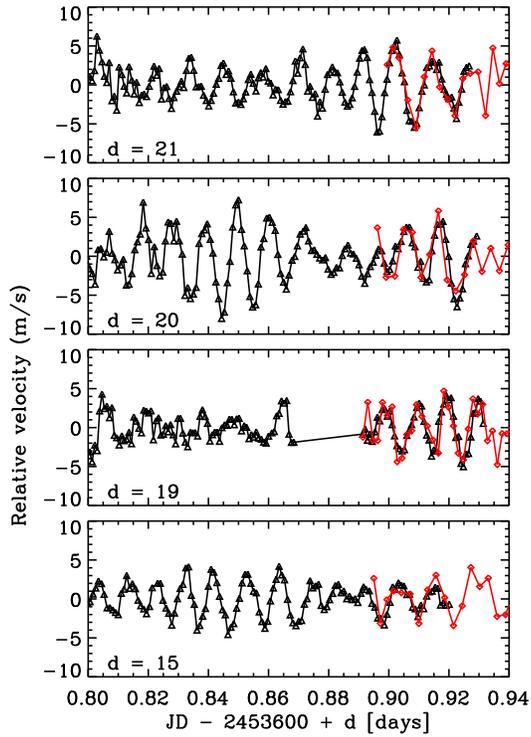}
\caption[]{\label{fig.overlap} Close-ups of the time series in
Fig.~\ref{fig.series} for the four segments during which both telescopes
were observing simultaneously.  Black triangles are from HARPS and red
diamonds are from UCLES\@.}
\end{figure}

The resulting velocities are shown in Fig.~\ref{fig.series}.  As can be
seen, the weather was very good in Chile but only moderately good in
Australia (we were allocated 7 nights with HARPS and 12 with UCLES).  The
seeing was also much better in Chile.  Most of the scatter in the
velocities is due to oscillations, but there are also slow variations and
night-to-night variations in both series that we ascribe to instrumental
effects.  Figure~\ref{fig.overlap} shows close-up views during the four
brief periods in which both telescopes were observing simultaneously.  Note
that the velocity offsets between the two data sets were adjusted
separately in each segment, to compensate for differential drifts between
the two instruments.  After this was done, we see excellent agreement
between the two data sets.

Figure~\ref{fig.power-both} shows the power spectra of the two time series.
The signal from stellar oscillations appears as the broad excess of power
centered at 1\,mHz.  As usual, we have used the measurement uncertainties,
$\sigma_i$, as weights in calculating these power spectra (according to
$w_i = 1/\sigma_i^2$).  In the case of HARPS, these uncertainties were
provided by the data processing pipeline.  For UCLES, they were estimated
from the scatter in the residuals during the measurement, as described by
\citet{BMW96}.

The peak in the HARPS power spectrum at 3070\,\muHz{} (and another at twice
this frequency, not shown here) is due to a periodic error in the guiding
system, as previously noted by \citet{C+E2006} and \citet{BazBK2007}.
Fortunately, this signal is restricted to a fairly narrow band of
frequencies that lies well above the oscillations of \bhyi, so it does not
compromise the data.

In analysing the data, we have followed basically the same method that we
used for \acena{} \citep{BBK2004,BKB2004}, \acenb{} \citep{KBB2005} and
\nuind{} \citep{BBC2006}.  Our initial goal was to adjust the weights in
order to minimize the noise level in the Fourier spectrum
(\S\,\ref{sec.noise-opt}).  Having done this, we then made further
adjustments with the aim of reducing the sidelobes in the spectral window
(\S\,\ref{sec.sidelobe-opt}).

\subsection{Optimizing for Signal-to-Noise}     \label{sec.noise-opt}

\begin{figure*}
\epsscale{0.9}
\plotone{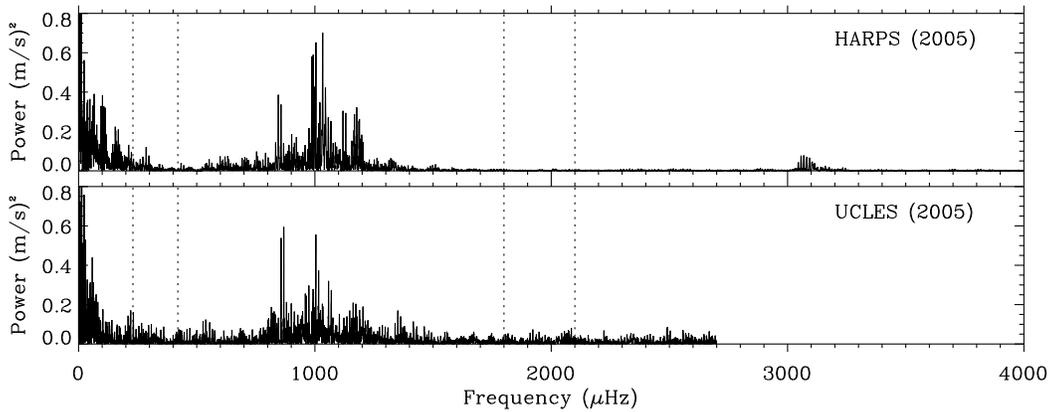}
\caption[]{\label{fig.power-both} Power spectra of \bhyi{} from the two
instruments, based on the data as it emerged from the pipeline (using the
uncertainties as weights).  The pairs of dotted lines mark the two regions
that were use to measure the noise level. }
\end{figure*}

We have chosen to measure the noise in the amplitude spectrum, $\sigma_{\rm
amp}$, in two frequency bands on either side of the oscillation signal:
230--420\,\muHz{} and 1800--2100\,\muHz, as indicated by the dotted lines
in Fig.~\ref{fig.power-both}.  We averaged these using the geometric mean
(since instrumental noise varies as an inverse power of frequency).  Using
this criterion the power spectra in Fig.~\ref{fig.power-both} have noise
levels (in amplitude) of 6.1\,\cms{} for HARPS and 11\,\cms{} for UCLES\@.
These values imply a noise per minute of observing time, before any further
optimization, of 1.9\,\ms{} for HARPS and 3.7\,\ms{} for UCLES\@.  The
difference is due to a combination of factors, primarily the observing duty
cycle and sky conditions.

The first step in optimizing the weights was to modify some of them to
account for a small fraction of bad data points, in the same way that we
have done for other stars (see \citealt{BBK2004} for details).  In brief,
this involved (i)~cleaning from the time series all power at low
frequencies (below 200\,\muHz), as well as all power from oscillations
(500--1400\,\muHz); and (ii)~searching these residuals for points that
deviated from zero by more than would be expected from their uncertainties.
We found that 78 data points from HARPS (2.8\%) and 25 from UCLES (2.1\%)
needed to be significantly down-weighted.  

The next step in reducing the noise involved ensuring that the
uncertainties we are using to calculate the weights reflect the actual
scatter in the data.  By this, we mean that the estimates of $\sigma_i$
should be consistent with the noise level determined from the amplitude
spectrum, which means they should satisfy equation~(3) of \citet{BBK2004}:
\begin{equation}
      \sigma_{\rm amp}^2 \sum_{i=1}^{N} \sigma_i^{-2}  = \pi. 
        \label{eq.condition}
\end{equation}
We checked this on a night-by-night basis for both instruments, measuring
$\sigma_{\rm amp}$ in the way described above.  The
results showed that the uncertainties $\sigma_i$ for HARPS should be
multiplied by a factor that ranged from 2.6 to 4.3, while for UCLES the
factor ranged from 0.57 to 0.73.  The variations in this factor from night
to night presumably reflect changes in the instrumental stability.  Note
that the factor for UCLES can be compared with the value of 0.87 that we
estimated for \acena{} \citep{BBK2004}, which was however determined from
measuring the noise at much higher frequencies and for the run as a whole.
For HARPS, the large discrepancy between the uncertainties estimated by the
pipeline and those required to explain the scatter in the data has been
noted previously, at levels consistent with our results
\citep{BBS2005,C+E2006,BazBK2007}.

With the velocity uncertainties corrected in this way, we found that HARPS
gave a mean precision per spectrum on \bhyi{} of about 1.4\,\ms{} and UCLES
gave 1.7\,\ms{} (with a spread during the run of about 20\%).  However,
HARPS was able to record spectra at about twice the rate of UCLES, thanks
to the better atmospheric conditions (which allowed shorter exposure times)
and the faster CCD readout.

\begin{figure*}
\epsscale{0.9}
\plotone{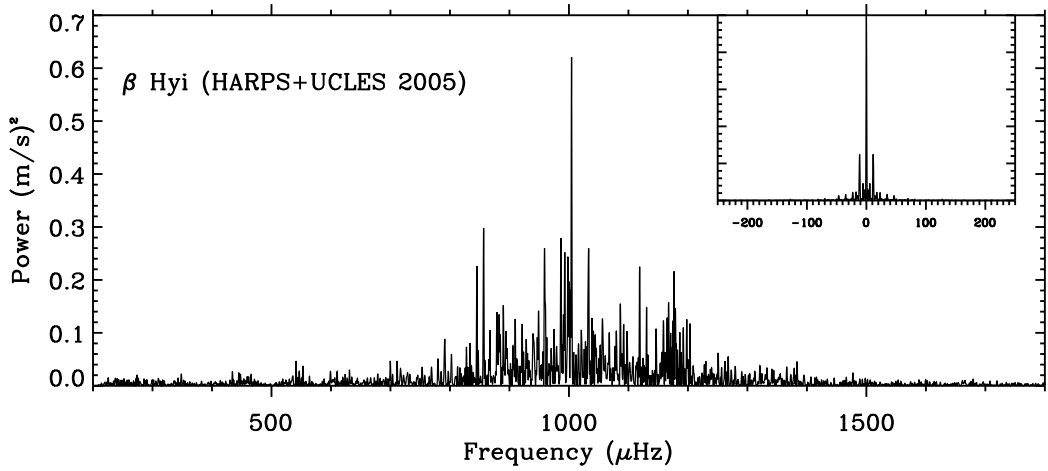}
\caption[]{\label{fig.power} Power spectrum of \bhyi{} after combining the
  data to optimize the signal-to-noise.  The inset shows the spectral
  window. }
\end{figure*}

The power spectrum of the combined time series is shown in
Fig.~\ref{fig.power} and the noise level is 4.2\,\cms{} in amplitude.  We
refer to this as the noise-optimized power spectrum.  Note that the time
series has been high-pass filtered in order to account for the varying
offsets between the two data sets and to prevent power from slow variations
from leaking into the oscillation signal.

\subsection{Optimizing for sidelobes}   \label{sec.sidelobe-opt}

\begin{figure}
\epsscale{0.45}
\plotone{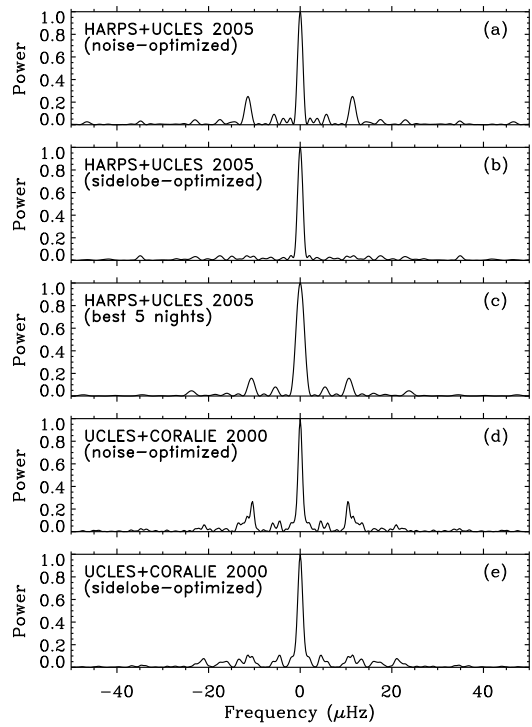}
\caption[]{\label{fig.windows} Spectral windows for the 2005 and 2000 data,
  using different weighting schemes (see text).  }
\end{figure}

The inset in Fig.~\ref{fig.power} shows the spectral window (the response
to a single pure sinusoid) and we see sidelobes at $\pm$ one cycle per day
($11.6$\,\muHz) that are moderately strong (25\% in power).  A close-up is
shown in Fig.~\ref{fig.windows}a.  These sidelobes occur despite there
being relatively few gaps in the observing window because of the higher
rate of data collection with HARPS as compared to UCLES\@.

As for our analysis of \acena{} and B, we have also generated a power
spectrum in which the weights were adjusted on a night-by-night basis in
order to minimize the sidelobes.  This involved giving greater weight to
the UCLES data (multiplying the weights by a factor that was typically
about 10) and resulted in a spectral window with sidelobes reduced to only
4.1\% in power (see Fig.~\ref{fig.windows}b).  The trade-off is an increase
in the noise level, which rose to 6.2\,\cms{} in amplitude.

Finally, we also calculated a power spectrum based on the best five nights
from both sites (${\rm JD} - 2453600 = 17.5$ to 22.5; see
Fig.~\ref{fig.series}).  In this case, the weights given to the five nights
of UCLES data were increased by a single factor of 7.2 to lower the
sidelobes, whose resulting height was 15.7\% in power (see
Fig.~\ref{fig.windows}c).  The noise in this amplitude spectrum was
6.2\,\cms{}.  This power spectrum has similar noise to the
sidelobe-optimized version and higher sidelobes, but it has the important
property of covering a shorter period of time.  We expect the modes in
\bhyi{} to have lifetimes of a few days, and so the best signal-to-noise
will be achieved in a time series that is not too much longer than this.
Once the observing time greatly exceeds the mode lifetime, the oscillation
modes in the power spectrum become resolved (eventually splitting into a
cluster of peaks under a Lorentzian envelope) and the signal-to-noise of
the peaks no longer improves with increased observing time.

\section{Re-analysis of the 2000 observations}    \label{sec.old}

\begin{figure*}
\epsscale{0.9}
\plotone{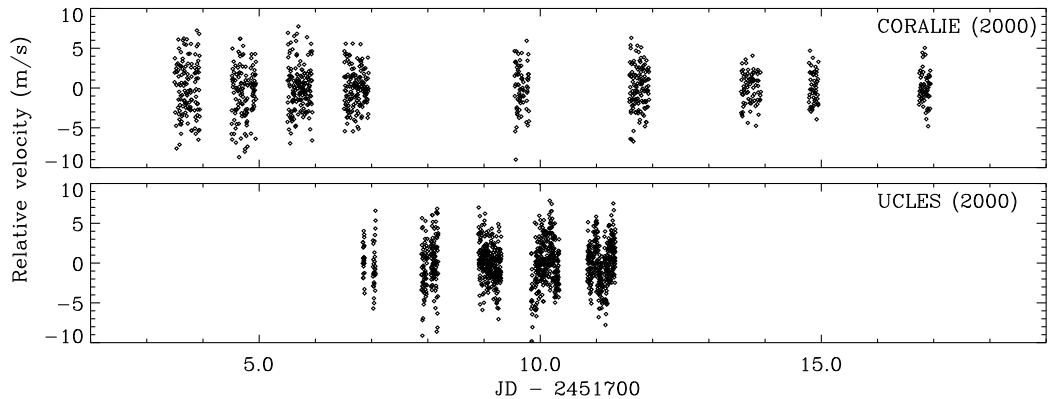}
\caption[]{\label{fig.series2000} Time series of velocity measurements of
\bhyi{} made in 2000 June with UCLES and CORALIE\@.  }
\end{figure*}

To allow us to detect as many oscillation modes as possible in \bhyi, our
analysis included our 2000 June observations.  These data consist of five
nights with UCLES \citep{BBK2001} and 14 nights with the CORALIE
spectrograph in Chile \citep{CBK2001}, with the latter being affected by
bad weather.  The two time series are shown in Fig.~\ref{fig.series2000}.

The UCLES \'echelle spectra have been completely reprocessed, using new
packages for the raw reduction and velocity extraction, resulting in an
improvement in precision at high frequencies of 13\%.  We have not
re-analyzed the CORALIE spectra and so these velocities are the same as
already published \citep{CBK2001}.

\begin{figure*}
\epsscale{0.9}
\plotone{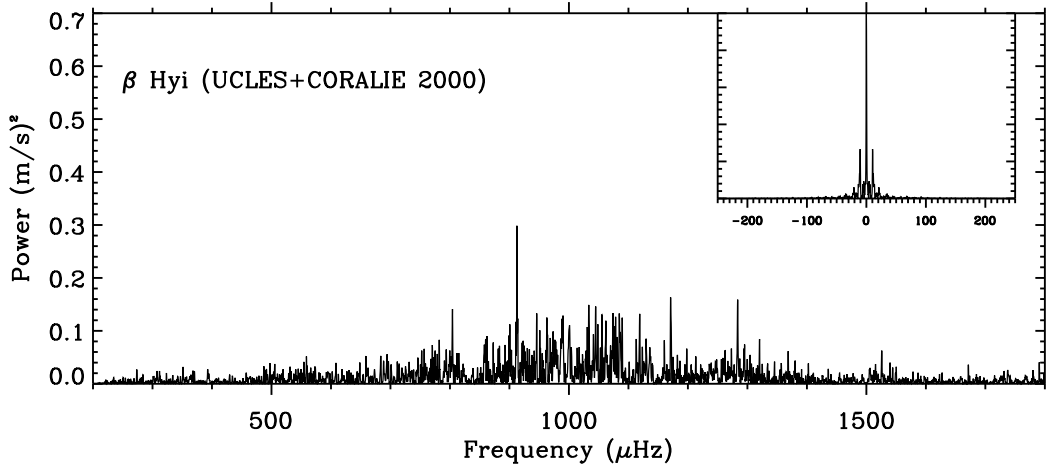}
\caption[]{\label{fig.power2000} Power spectrum of \bhyi{} from the 2000
observations with UCLES and CORALIE\@.  }
\end{figure*}

We subjected both 2000 velocity time series to the same analysis that is
described above.  The combined noise-optimized power spectrum is shown in
Fig.~\ref{fig.power2000}.  The noise averaged over our defined frequency
ranges is 5.9\,\cms{} in amplitude.  The spectral window is shown in the
inset and also in close-up in Fig.~\ref{fig.windows}d.  The
sidelobe-optimized version of these data has a noise level of 8.0\,\cms{}
and its spectral window is shown in Fig.~\ref{fig.windows}e.

\section{Oscillation frequencies}    \label{sec.freqs}

Mode frequencies for low-degree p-mode oscillations in main-sequence stars
are well approximated by a regular series of peaks, with frequencies given
by the following asymptotic relation:
\begin{equation}
  \nu_{n,l} = \Dnu (n + \half l + \epsilon) - l(l+1) D_0.
        \label{eq.asymptotic}
\end{equation}
Here $n$ (the radial order) and $l$ (the angular degree) are integers,
$\Dnu$ (the large separation) depends on the sound travel time across the
whole star, $D_0$ is sensitive to the sound speed near the core and
$\epsilon$ is sensitive to the surface layers.  See \citet{ChD2004} for a
recent review of the theory of solar-like oscillations.

A subgiant such as \bhyi{} is expected to show substantial deviations from
the regular comb-like structure described by
equation~(\ref{eq.asymptotic}).  This is because some mode frequencies,
particularly those with $\ell=1$, are shifted by avoided crossings with
gravity modes in the stellar core (also called `mode bumping'; see
\citealt{ASW77}).  These shifted modes are known as `mixed modes' because
they have p-mode character near the surface but g-mode character in the
deep interior.  Theoretical models of \bhyi{} indeed predict these mixed
modes \citep{DiMChDP2003,F+M2003}, and we must keep this in mind when
attempting to identify oscillation modes in the power spectrum.  The mixed
modes are rich in information because they probe the stellar core and are
very sensitive to age, but they complicate greatly the task of mode
identification.

We have extracted oscillation frequencies from the time series using the
standard procedure of iterative sine-wave fitting.  \newnew{Each step of
the iteration involves finding the strongest peak in the power spectrum and
subtracting the corresponding sinusoid from the time series.  At each step,
the frequencies, amplitudes and phases of all extracted peaks were adjusted
simultaneously to give the best fit.}  In analysing these frequencies, the
first step was to determine the large separation.  We did this by examining
the pair-wise differences between frequencies extracted from the
noise-optimized 2005 power spectrum.  We included all peaks with $S/N \ge
4$ (46 peaks) and found the strongest signals at 29.1 and 57.2\,\muHz.  We
identify these as corresponding to $\Dnu/2$ and $\Dnu$, respectively,
consistent with the values published from the 2000 observations
\citep{BBK2001,CBK2001}.

\begin{figure*}
\epsscale{0.45}
\plotone{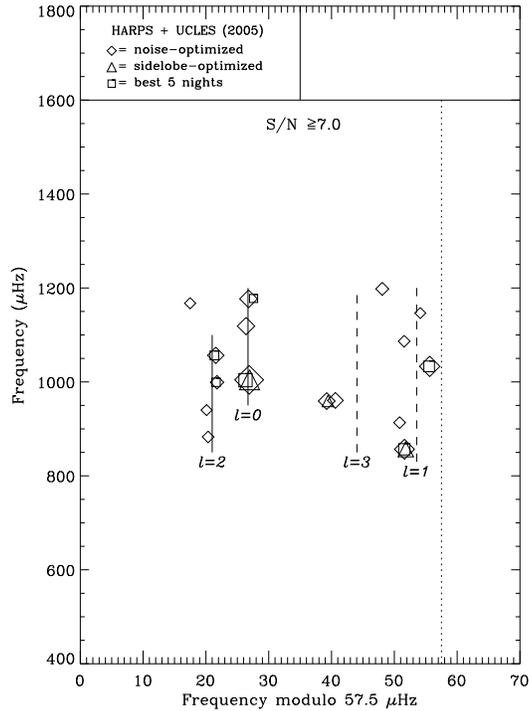}
\caption[]{\label{fig.echelle7} \'Echelle diagram of frequencies extracted
from the 2005 observations.  Only the strongest peaks are shown ($S/N \ge
7$).  Different symbols identify the three weighting schemes (see
\S\,\ref{sec.noise-opt} and \S\,\ref{sec.sidelobe-opt}) and the symbol size
is proportional to the S/N\@. The solid lines mark our identification of
$l=0$ and $l=2$, and these were used to calculate the positions for $l=1$
and $l=3$ (dashed lines), using the asymptotic relation
(equation~\ref{eq.asymptotic}). }
\end{figure*}

With the large separation established, we next sought to identify modes in
the \'echelle diagram.  Figure~\ref{fig.echelle7} shows only the very
strongest peaks in the 2005 data ($S/N \ge 7$).  Different symbols are used
to show the three weighting schemes and the symbol size is proportional to
the S/N\@.  We can immediately identify the ridges corresponding to modes
with $l=0$ and $l=2$, both of which are expected to be unaffected by
avoided crossings.  The positions of these two ridges, marked by the
vertical solid lines, allow us to calculate the expected positions of modes
with $l=1$ and $l=3$, based on the asymptotic relation
(equation~\ref{eq.asymptotic}).  These are shown by dashed lines in the
figure.  A few of the peaks in that region of the diagram fall close to the
$l=1$ line, but most do not.  We are clearly seeing several mixed modes in
\bhyi.

Deciding which peaks correspond to genuine modes is always a subjective
process.  There is a trade-off between the desire to find as many
oscillation modes as possible and the risk of identifying noise peaks as
genuine.  There is also the danger, especially for the weaker peaks, that
the sidelobe will be stronger than the real peak because of interference
with the noise, leading to an aliasing error of $\pm11.6\,\muHz$.  And for
an evolved star like \bhyi, we have the added complication of mixed modes.

\begin{figure*}
\epsscale{0.45}
\plotone{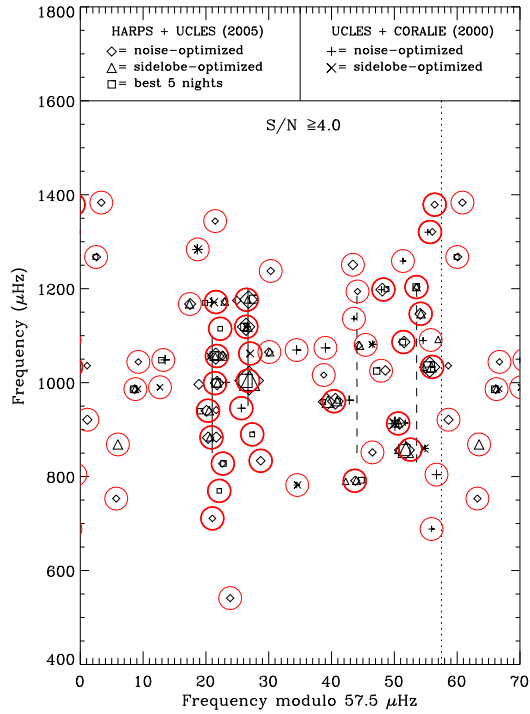}
\caption[]{\label{fig.echelle4} Same as Fig.~\ref{fig.echelle7} but also
  including the 2000 observations (see \S\,\ref{sec.old}) and with all
  peaks having $S/N\ge4$.  The red circles show the frequencies listed in
  Table~\ref{tab.freqs}, with thicker lines denoting modes for which we are
  confident of the identification. }
\end{figure*}

In the case of \bhyi, we have the advantage of two independent data sets
taken five years apart.  The 2005 data have extremely low noise and a very
good window function.  The 2000 data have higher noise and more gaps, but
the sidelobe-optimized weighting still gives a good spectral window.  In
Fig.~\ref{fig.echelle4} we show all peaks extracted from the time series
that have $S/N \ge 4$.  At this level, very few of the peaks should be due
to noise.  We again used different symbols to identify the different
weighting schemes, as shown in the figure legend.

The circles in Fig.~\ref{fig.echelle4} show the final frequencies, which
are listed in Table~\ref{tab.freqs}.  Given the effects of noise and finite
mode lifetime, we do not expect perfect agreement between different
measurements of the same mode.  In cases where a peak was detected in more
than one weighting scheme, we have averaged the frequency measurements.  In
those cases, the S/N in the table refers to the maximum among the
contributors.  

\begin{figure}
\epsscale{0.45}
\plotone{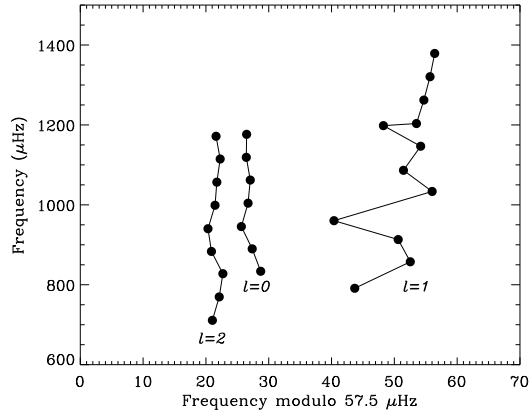}
\caption[]{\label{fig.echelle-final} \'Echelle diagram of the 28 oscillation
frequencies in \bhyi{} of which we are confident.  The frequencies are
given in the upper part of Table~\ref{tab.freqs}. }
\end{figure}

The 28 frequencies above the line are those for which we are confident of
the identification.  These are also shown in Fig.~\ref{fig.echelle-final},
where we see a pattern of mixed modes that is strikingly similar to that
calculated from models by \citet{DiMChDP2003}.  \newnew{Note that the $n$
values in Table~\ref{tab.freqs} were determined on the assumption that
$\epsilon$ falls in the range $1\le \epsilon \le 2$, as it does in the
Sun.}  The 29 frequencies below the line in the table will include some
genuine modes, some sidelobes that need to be shifted by $\pm11.6\,\muHz$,
and some noise peaks.  We list them here for completeness, to allow
comparison with oscillation models of \bhyi.

The uncertainties in the mode frequencies are shown in parentheses in
Table~\ref{tab.freqs}.  These depend on the S/N ratio of the peak and were
calculated from the simulations described in \S\,\ref{sec.lifetime},
assuming the value of the mode lifetime derived in that section.  Of
course, these frequency uncertainties are based on the assumption that the
corresponding peaks in the power spectrum are caused by a genuine
oscillation modes and are not noise peaks or aliases.

We have looked for a systematic offset between mode frequencies of \bhyi{}
from the 2005 and 2000 observations.  Such an offset could indicate
variations during a stellar activity cycle, as has tentatively been
suggested for \acena{} \citep{FCE2006}.  We find that the 2005 frequencies
are lower than the 2000 frequencies by an average amount of
$0.1\pm0.4$\,\muHz, which is consistent with zero.  \newnew{Finally, of the
two mode identifications suggested by \citet{BBK2001}, we can now identify
solution B as being closest to the correct one.}

\subsection{Frequencies from a Bayesian treatment}\label{sec.bayesian}

We have recently described an alternative approach to extracting
frequencies from a time series, which we applied to the star \nuind, using
Bayesian methods rather than Fourier analysis \citep{BBC2006,BBK2007}.  We
have applied a similar technique to \bhyi, using the noise-optimized 2005
time series.  This analysis differed in one important respect from that on
\nuind: we did not make any assumptions about the distribution of
frequencies.

\begin{figure*}
\epsscale{0.45}
\plotone{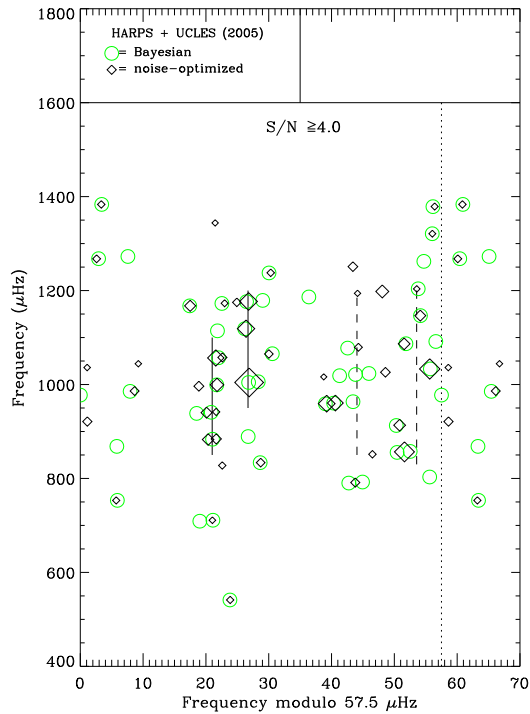}
\caption[]{\label{fig.echelle.brendon} A comparison between frequencies
  present in the noise-optimized 2005 time series found using traditional
  iterative sine-wave fitting (diamonds; same as in
  Fig.~\ref{fig.echelle4}) and the most probable peaks found using a
  Bayesian method (green circles; see \S\,\ref{sec.bayesian}).}
\end{figure*}

The results are shown in Fig.~\ref{fig.echelle.brendon}, where the diamonds
are the same as in Fig.~\ref{fig.echelle4} and the circles are the most
probable frequencies found by the Bayesian method.  In most cases there is
excellent agreement between the two analyses.  There is a peak at
1262.20\,\muHz, which lies exactly on the $l=1$ ridge, that only appears in
Bayesian analysis.  In fact, this peak was found with S/N = 3.0 by the
traditional method and we have shown it as a genuine mode in
Table~\ref{tab.freqs} and Fig.~\ref{fig.echelle-final}.

\subsection{Frequency separations and inferred stellar parameters}

The large separation \Dnu{} is a function of the mode degree,~$l$, and also
varies with frequency.  We have determined it for both $l=0$ and~2 by
fitting to the modes identified in Table~\ref{tab.freqs} and shown in
Fig.~\ref{fig.echelle-final}.  The values at 1\,mHz are given in
Table~\ref{tab.params}.  We also give the small separation $\delta\nu_{02}$
between $l=0$ and~2, and the implied values for the parameters $D_0$
and~$\epsilon$.

A detailed comparison of the oscillation frequencies of \bhyi{} with
theoretical models is beyond the scope of this paper.  Here, we restrict
ourselves to using the large separation to estimate the mean density of the
star.  To a good approximation, the large separation is proportional to the
square root of the mean stellar density:
\begin{equation}
  \frac{\Dnu}{\Dnu_\sun} =  \sqrt{\frac{\bar{\rho}}{\bar{\rho}_\sun}}.\label{eq.rho}
\end{equation}
However, it is well-known that models -- even for the Sun -- are not yet
good enough to reproduce the observed frequencies due to improper modelling
of the surface layers \citep{ChDDL88}.  There is a systematic offset
between observed and computed frequencies that increases with frequency,
and hence leads to an incorrect prediction for the large separation.

We have examined this surface effect in detail for models of the Sun and
\bhyi{} (Kjeldsen \& Bedding, in prep.) and found that the correction term
scales in such a way that, provided that $\Delta\nu$ is measured at the
peak of the oscillation envelope, equation~(\ref{eq.rho}) remains an
excellent approximation.  We have therefore used equation~(\ref{eq.rho}) to
estimate the mean density of \bhyi{} from our measurement of $\Dnu_0$.  We
used a value of $134.81 \pm 0.09$\,\muHz{} for the large separation of
radial modes in the Sun, which we obtained by fitting in the range
$n=17$--25 to the frequencies measured by \citet{LBB97} using the GOLF
instrument on the SOHO spacecraft.  The resulting mean density for \bhyi{}
has an uncertainty of only 0.6\% and is given in Table~\ref{tab.params}.


\section{Oscillation amplitudes}  \label{sec.amp}

The amplitudes of individual modes are affected by the stochastic nature of
the excitation and damping.  To measure the oscillation amplitude of
\bhyi{} in a way that is independent of these effects, we have followed the
method introduced by \citet{KBB2005}.  In brief, this involves the
following steps: (i)~smoothing the power spectrum heavily to produce a
single hump of excess power that is insensitive to the fact that the
oscillation spectrum has discrete peaks; (ii)~converting to power density
by multiplying by the effective length of the observing run (which we
calculated from the area under the spectral window in power); (iii)~fitting
and subtracting the background noise; and (vi)~multiplying by
$\Delta\nu/3.0$ and taking the square root, in order to convert to
amplitude per oscillation mode.  For more details, see \citet{KBB2005}.

\begin{figure}
\epsscale{0.45}
\plotone{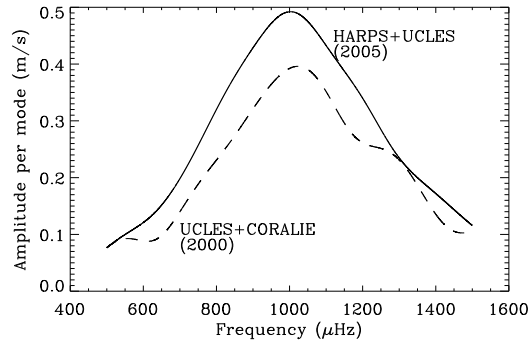}
\caption[]{\label{fig.amp} Amplitude of oscillations in \bhyi{} from the
2000 and 2005 observations (see \S\,\ref{sec.amp}).  }
\end{figure}

The result is shown in Fig.~\ref{fig.amp} for both the 2005 and 2000
observations.  The difference between the two runs, which were made five
years apart, is not surprising when we recall that the solar amplitude,
when measured in the same way, shows similar variations from week to week.
The peak amplitude per mode in \bhyi{} is 40--50\,\cms, which occurs at a
frequency of $\nu_{\rm max}=1.0$\,mHz.  This value of $\nu_{\rm max}$ is
consistent with that expected from scaling the acoustic cutoff frequency of
the Sun \citep{BGN91,K+B95}.  The observed peak amplitude is about 2.5
times the solar value, when the latter is measured using stellar techniques
(Kjeldsen et al., in prep.).  Given the uncertainties, this is consistent
with the values of 3.2 expected from the $L/M$ scaling proposed by
\citet{K+B95} and also with 2.2 from the $(L/M)^{0.7}$ scaling proposed by
\citet{SGA2005}.

\section{Mode lifetimes}\label{sec.lifetime}

We can estimate the lifetime of solar-like oscillations from the scatter of
the observed frequencies about the ridges in the \'echelle diagram (see
\citealt{KBB2005}).  For \bhyi{} we used only the $l=0$ and~$2$ modes,
since these are apparently unaffected by avoided crossing.  In addition, we
examined the differences between the 2000 and 2005 data sets for those
modes that were detected in both (assuming that any differences due to a
stellar activity cycle are negligible; see \S\,\ref{sec.freqs}).  We found
the mean scatter per mode to be $1.25 \pm 0.15$\,\muHz.  Following the
method described by \citet{KBB2005}, this frequency scatter was calibrated
using simulations having a range of mode lifetimes and with the observed
S/N, window function and weights.  We assumed that any rotational splitting
could be neglected, which is certainly reasonable given the very low
measured rotational velocity of the star of 2--5\,k\ms
\citep{D+N90,R+S2003,SPdS2004}.

\begin{figure}
\epsscale{0.45}
\plotone{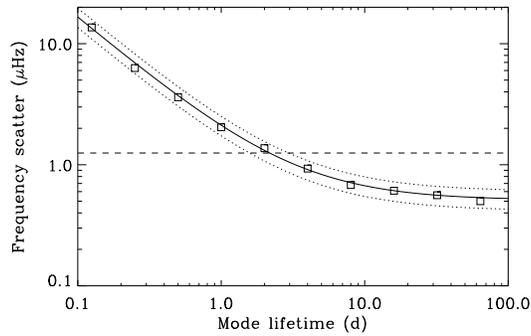}
\caption[]{\label{fig.lifetime.freq} Calibration of the mode lifetime from
measurements of the frequency scatter.  Each point shows the mean of 300
simulations, each measuring the frequency scatter for a given mode
lifetime.  The solid curve is a fit to these points and the dotted curves
show the 1-$\sigma$ variations in the simulations.  The horizontal dashed
line shows the observed scatter in the frequencies, from which we can read
off the mode lifetime of \bhyi{} (see \S\,\ref{sec.lifetime}).

}
\end{figure}

We carried out the simulations for ten different values of the mode
lifetime and the results are shown in Figure~\ref{fig.lifetime.freq}, where
each point represents the mean of 300 simulations.  The solid curve
represents a fit to these points and the two dotted curves on either side
reflect the 1-$\sigma$ variations in the simulations.  The horizontal
dashed line shows the observed scatter in the frequencies, from which we
can read off a value for the mode lifetime of $2.18^{+0.77}_{-0.60}$\,d.

As a check, we also used a new and completely different method to estimate
the mode lifetime.  This involved examining the region of excess power in
the Fourier spectrum and measuring the ratio between the mean in the power
spectrum and the square of the mean amplitude.  This ratio, $R = \langle
A^2\rangle / \langle A\rangle^2$, measures the `peakiness' of the power
spectrum.  For purely Gaussian noise, $R$ will have a value of $4/\pi =
1.27$ (see equation~(A2) of \citealt{K+B95}), while for a spectrum composed
of many strong narrow peaks it will have a much higher value.  For example,
the observed value of $R$ in the noise-optimized spectrum from 2005 is 1.42
(measured in the range 700--1300\,\muHz).  Note that the ratio of
power-to-squared-amplitude is only a sensitive indicator of mode lifetime
in the case where the noise level in the spectrum is very low, which is
true for our 2005 observations of \bhyi.

We have again calibrated the observed values of $R$ by comparing with
simulations with a range of mode lifetimes.  The only additional
information that we required, beyond repeating the assumption of negligible
rotation, was the number of oscillation modes in the spectrum.  This is
easily calculated from the large separation, provided we assume that there
is exactly one mode for each pair of $l$ and $n$.  For extremely evolved
stars in which the spectrum of g-modes is very dense, we would expect more
than one mixed mode per order, but for \bhyi{} this does not appear to be
the case \citep{DiMChDP2003,F+M2003}.  It is therefore reasonable to assume
that, despite the presence of mixed modes, there is still on average only
one mode per order for each value of the degree~$l$.  

\begin{figure}
\epsscale{0.45}
\plotone{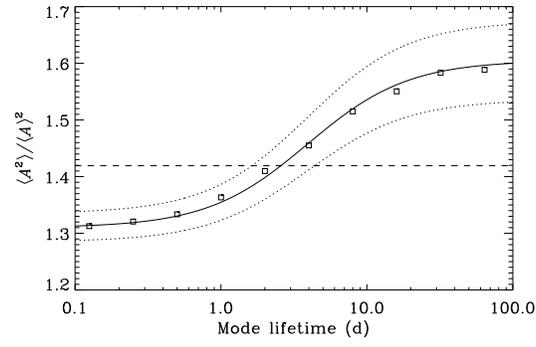}
\caption[]{\label{fig.lifetime.amp} Calibration of the mode lifetime from
measurements of the ratio of power-to-squared-amplitude in the
noise-optimized power spectrum.  Each point shows the mean of 300
simulations, each measuring the frequency scatter for a given mode
lifetime.  The solid curve is a fit to these points and the dotted curves
show the 1-$\sigma$ variations in the simulations.  The horizontal dashed
line shows the observed value of the ratio (see \S\,\ref{sec.lifetime}).  }
\end{figure}

We again carried out the simulations for ten different values of the mode
lifetime.  The results for the noise-optimized spectrum from 2005 are shown
in Figure~\ref{fig.lifetime.amp}, where each point represents the mean of
300 simulations.  The solid curve represents a fit to these points and the
two dotted curves on either side reflect the 1-$\sigma$ variations in the
simulations.  The horizontal dashed line shows the observed value of $R$,
which allows us to determine the mode lifetime.  We did the same thing for
the other two weighting schemes for the 2005 data (sidelobe-optimized and
best five nights).  The mean of the three schemes gave a mode lifetime of
$2.57^{+1.37}_{-0.82}$\,d.

These two estimates are independent in the sense that they rely on two
different properties of the power spectrum.  The first measures the
positions of the peaks and the second measures their amplitudes.  We are
therefore justified in combining the two estimates for the mode lifetime,
and the final value is listed in Table~\ref{tab.params}.  This value,
combined with the amplitudes estimated in \S\,\ref{sec.amp}, should allow a
comparison with theoretical models of the excitation of solar-like
oscillations \citep{SGT2007,Hou2006}.

\section{Conclusions}

We have presented dual-site velocity observations of \bhyi{} that confirm
the presence of solar-like oscillations.  These data, combined with those
we obtained in 2000, allowed us to identify 28 oscillation modes.  The
large frequency separation allowed us to infer a very precise value for
the mean density of \bhyi.  We also measured the amplitudes and lifetimes of the
oscillations.  Finally, the frequencies show the clear signature of mixed
$l=1$ modes which, after comparison with models of this star, should allow
a sensitive measure of its age.

We would be happy to make the data presented in this paper available on
request.

\acknowledgments

We thank Geoff Marcy for useful advice and enthusiastic support and Dennis
Stello for comments on the manuscript.  This work was supported financially
by the Australian Research Council, the Swiss National Science Foundation,
the Portuguese FCT and POCI2010 (POCI/CTE-AST/57610/2004) with funds from
the European programme FEDER and the Danish Natural Science Research
Council.  We further acknowledge support by NSF grant AST-9988087 (RPB) and
by SUN Microsystems.

\clearpage

\begin{table*}
\begin{center}
\caption{\label{tab.freqs}Oscillation frequencies in \bhyi}
\begin{tabular}{rrrl}
\tableline
\noalign{\smallskip}
\tableline
\noalign{\smallskip}
\multicolumn{1}{c}{$\nu$}    & \multicolumn{1}{c}{$\nu \bmod \Delta\nu$} &     &         \\
(\muHz)  & \multicolumn{1}{c}{(\muHz)}               & S/N & \sc Mode ID \\
\noalign{\smallskip}
\tableline
\noalign{\smallskip}
  711.03 (1.49) & 21.0 & 4.3 & $l=  2$, $n=10$ \\
  769.62 (1.59) & 22.1 & 4.1 & $l=  2$, $n=11$ \\
  791.19 (1.06) & 43.7 & 6.0 & mixed \\      
  827.70 (1.19) & 22.7 & 5.2 & $l=  2$, $n=12$ \\
  833.72 (1.10) & 28.7 & 5.7 & $l=  0$, $n=13$ \\
  857.54 (0.86) & 52.5 &13.0 & $l=  1$, $n=13$ \\
  883.38 (0.94) & 20.9 & 7.5 & $l=  2$, $n=13$ \\
  889.87 (1.45) & 27.4 & 4.4 & $l=  0$, $n=14$ \\
  913.11 (0.89) & 50.6 & 9.3 & $l=  1$, $n=14$ \\
  940.33 (0.95) & 20.3 & 7.2 & $l=  2$, $n=14$ \\
  945.64 (1.13) & 25.6 & 5.5 & $l=  0$, $n=15$ \\
  960.39 (0.87) & 40.4 &10.9 & mixed \\      
  998.95 (0.90) & 21.5 & 8.8 & $l=  2$, $n=15$ \\
 1004.21 (0.85) & 26.7 &18.8 & $l=  0$, $n=16$ \\
 1033.52 (0.86) & 56.0 &13.3 & $l=  1$, $n=16$ \\
 1056.74 (0.87) & 21.7 &10.7 & $l=  2$, $n=16$ \\
 1062.06 (1.27) & 27.1 & 4.9 & $l=  0$, $n=17$ \\
 1086.45 (0.93) & 51.4 & 7.6 & $l=  1$, $n=17$ \\
 1114.77 (1.49) & 22.3 & 4.3 & $l=  2$, $n=17$ \\
 1118.93 (0.87) & 26.4 &11.3 & $l=  0$, $n=18$ \\
 1146.69 (0.95) & 54.2 & 7.2 & $l=  1$, $n=18$ \\
 1171.61 (0.97) & 21.6 & 7.0 & $l=  2$, $n=18$ \\
 1176.48 (0.87) & 26.5 &11.6 & $l=  0$, $n=19$ \\
 1198.26 (0.90) & 48.3 & 8.5 & mixed \\      
 1203.52 (1.04) & 53.5 & 6.1 & $l=  1$, $n=19$ \\
 1262.20 (2.66) & 54.7 & 3.0 & $l=  1$, $n=20$ \\
 1320.68 (1.40) & 55.7 & 4.5 & $l=  1$, $n=21$ \\
 1378.92 (1.33) & 56.4 & 4.7 & $l=  1$, $n=22$ \\
\noalign{\smallskip}
\tableline      
\noalign{\smallskip}
  541.36 (1.33) & 23.9 & 4.7\\
  688.43 (1.59) & 55.9 & 4.1\\
  753.22 (1.33) &  5.7 & 4.7\\
  782.04 (1.37) & 34.5 & 4.6\\
  804.21 (1.02) & 56.7 & 6.3\\
  851.51 (1.27) & 46.5 & 4.9\\
  868.49 (1.03) &  6.0 & 6.2\\
  921.12 (1.04) &  1.1 & 6.1\\
  986.27 (1.04) &  8.8 & 6.1\\
  990.19 (1.59) & 12.7 & 4.1\\
 1016.26 (1.59) & 38.8 & 4.1\\
 1025.39 (1.02) & 47.9 & 6.3\\
 1044.24 (1.54) &  9.2 & 4.2\\
 1048.21 (1.03) & 13.2 & 6.2\\
 1065.10 (1.11) & 30.1 & 5.6\\
 1069.44 (1.07) & 34.4 & 5.9\\
 1074.03 (1.02) & 39.0 & 6.3\\
 1080.44 (1.17) & 45.4 & 5.3\\
 1090.78 (1.45) & 55.8 & 4.4\\
 1136.06 (1.65) & 43.6 & 4.0\\
 1167.44 (0.95) & 17.4 & 7.2\\
 1194.14 (1.59) & 44.1 & 4.1\\
 1237.82 (1.27) & 30.3 & 4.9\\
 1250.90 (1.02) & 43.4 & 6.3\\
 1258.87 (1.49) & 51.4 & 4.3\\
 1267.51 (1.22) &  2.5 & 5.1\\
 1283.70 (0.99) & 18.7 & 6.7\\
 1343.97 (1.54) & 21.5 & 4.2\\
 1383.34 (1.27) &  3.3 & 4.9\\
\noalign{\smallskip}
\tableline

\end{tabular}

\end{center}
\end{table*}

\begin{table*}
\small
\caption{\label{tab.params} Parameters for \bhyi{}}
\begin{center}
\begin{tabular}{lc}
\tableline
\noalign{\smallskip}
\tableline
\noalign{\smallskip}
Parameter & Value \\
\noalign{\smallskip}
\tableline
\noalign{\smallskip}
$\Delta\nu_0$ at 1\,mHz (\muHz)          & $57.24 \pm 0.16$      \\
$\Delta\nu_2$ at 1\,mHz (\muHz)          & $57.52 \pm 0.10$      \\
$\delta\nu_{02}$ at 1\,mHz (\muHz)       & $5.32 \pm 0.45$       \\
$D_0$            at 1\,mHz (\muHz)       & $0.89 \pm 0.07$       \\
$\epsilon$       at 1\,mHz               & $1.55 \pm 0.05$       \\
$\bar{\rho}$ (g\,cm$^{-3}$)              & $0.2538 \pm 0.0015$  \\
mode lifetime (d)                        & $2.32^{+0.64}_{-0.51}$\\
\noalign{\smallskip}
\tableline
\end{tabular}
\end{center}
\end{table*}

\end{document}